\documentclass[twocolumn,showpacs,preprintnumbers,amsmath,amssymb]{revtex4}
\def\frac#1#2{{\textstyle{#1\over#2}}} 

\def\bra#1{\langle #1 |}
\def\ket#1{| #1\rangle}
\def\braket#1#2{\langle \, #1 \, | \, #2 \, \rangle}

\def\R{\hbox{\rm I \kern-5pt R}}

\def\ajou#1&#2(#3){\ \sl#1\bf#2\rm(19#3)}

\def\Tr{{\rm{Tr}}}

\begin{document}
\title{Cheat Sensitive Quantum Bit Commitment}
\author{ Lucien Hardy}
\affiliation{ The Perimeter Institute, 35 King Street North, Waterloo,
Ontario, Canada N2J 2W9}
\author{Adrian Kent}
\affiliation{Centre for Quantum Computation, Department of 
Applied Mathematics and
Theoretical Physics,\\ University of Cambridge,
Wilberforce Road, Cambridge CB3 0WA, United Kingdom}

\date{November 1999; revised February 2004}

\begin{abstract}
We define {\it cheat sensitive} cryptographic protocols between
mistrustful parties as protocols which guarantee that, if either
cheats, the other has some nonzero probability of detecting the
cheating. We describe an unconditionally secure cheat
sensitive non-relativistic bit commitment protocol which uses
quantum information to implement a task which is classically
impossible; we also describe a simple relativistic protocol.
\end{abstract}

\maketitle


The discovery of quantum cryptography\cite{wiesner}
and secure quantum key distribution\cite{BBef} has led to much
interest in understanding precisely which cryptographic tasks can
be guaranteed secure by physical principles. We propose here a new
class of cryptographic applications of quantum information: {\it
cheat sensitive} protocols between mistrustful parties.  Either party may be able to evade the
intended constraints on information transfer by deviating from these
protocols.  However, if they do, there is a non-zero probability
that the other will detect their cheating.  Cheat sensitivity is
potentially useful in any situation where the parties, though
mistrustful, have an ongoing relationship which they value more
than the potential gains from a few successful cheating attempts.  

We consider here cheat sensitive protocols for bit commitment
(BC), an important cryptographic primitive whose potential for
physically secure implementation has been extensively
investigated\cite{BBef,BCJL,brassardcrepeau,lochauprl,mayersprl,mayerstrouble,lochau,mayersone,bcms,kentrel,kentrelfinite,kentbccc}.
We first introduce BC and briefly review what is
currently known about physical implementations. 

Suppose $A$ and $B$ are in two different places and can send classical
messages or quantum states to one another.  In a classical BC
protocol, $A$ {\it commits} herself by giving $B$ information corresponding to an 
encryption of a bit (either 0 or 1), in such a way that she can
later decrypt or {\it unveil} it for him if she chooses.   It should be hard
(ideally, impossible) for $A$ to change the bit or
for $B$ to obtain any information about the bit unless and until $A$ chooses
to unveil it for him.  One possible implementation is for $A$ to
write the bit down on a piece of paper which she locks in a safe, 
and then send the locked safe to $B$.  To unveil the bit later, she sends 
$B$ the key to the safe.  One weakness of this
method is that $B$ might be able to open the safe unaided, for example by
picking the lock, or view its contents without opening it, for
example by magnetic resonance imaging.

In general, the commitment and unveiling follow a
prescribed protocol of information exchanges.
We distinguish {\it quantum}
protocols, which allow the exchange of quantum information, from {\it classical} 
protocols, which do not.  
We also distinguish {\it relativistic} protocols, which assume
the validity of special relativity and rely on the impossibility of
superluminal signalling, from {\it non-relativistic} protocols, which 
do not.   A BC protocol is {\it secure}, modulo certain
assumptions, if it includes a parameter which can be adjusted so
that the probabilities of $A$ being able to unveil a state significantly
different from the committed state and of $B$ being able to extract
significant information can simultaneously be made arbitrarily
small. It is {\it unconditionally secure} within a given physical
theory if the only assumption necessary is the validity of that
theory.  (Formal security definitions can be found in, e.g., 
Refs. \cite{mayersprl,kentbccc}.) 

Several quantum BC schemes have been proposed
\cite{BBef,BCJL,brassardcrepeau,salvail}. These schemes presently
offer good practical security, but in principle are insecure.
Indeed, Lo-Chau\cite{lochauprl,lochau} and 
Mayers\cite{mayersprl,mayerstrouble,mayersone} showed that no
non-relativistic quantum BC schemes can be perfectly secure
against both parties.  Mayers\cite{mayersprl,mayerstrouble,mayersone}
extended this to prove the impossibility of unconditionally 
secure non-relativistic quantum BC and to cover models in which
classical and quantum information are treated separately.  
Unconditionally secure
classical relativistic BC protocols exist\cite{kentrel,kentrelfinite}. 
It is conjectured that these protocols are also secure against
quantum attack.

The essential weakness in non-relativistic quantum BC protocols
highlighted by Lo-Chau and Mayers is that, whenever $B$ can
extract no (or little) information about the committed bit, $A$
can (or very probably can) undetectably alter the commitment from $0$ to
$1$ or vice versa.

A separate issue in quantum protocols is that
$A$ may commit an {\it improper} mixture of bit
states\cite{bcms,kentbccc}, in which the committed bit is
entangled with another state. This is not advantageous when BCs
are used in isolation --- for example, to record a secret
prediction --- but can be when they are used as subprotocols
for a larger task. A BC protocol which forces $A$ to
commit a fixed classical bit, $0$ or $1$, is said to implement bit
commitment with a certificate of classicality (BCCC). Secure BCCC
protocols based on reasonable quantum computational complexity
assumptions might be possible\cite{salvail,akfounding}, but no
unconditionally secure BCCC protocols exist\cite{kentbccc}.

Relativistic BC schemes\cite{kentrel,kentrelfinite} 
offer the prospect of practical unconditional security.  However,
they require the maintenance of separated sites and the continual
use of communication channels between commitment and unveiling.  For some
applications these constraints may be serious disadvantages.  Hence,
as unconditionally secure non-relativistic BC
is impossible, it is interesting to explore what can be achieved by
non-relativistic protocols. We show here that cheat sensitive BC
can be implemented non-relativistically.  We also give
a simple relativistic cheat sensitive BC protocol.

{\it Defining cheat sensitivity} \qquad A cheat
sensitive quantum BC protocol is a quantum BC
protocol in which, {\it assuming that the commitment will
eventually be unveiled}, $A$ cannot alter the
probabilities of her revealing a $0$ or $1$ after the commitment
without risking detection and $B$ cannot extract information
about the committed bit before the unveiling without risking
detection.  Note that by this definition the detection probabilities 
only need be nonzero to imply cheat sensitivity. 

{\it Protocols}  \qquad  If a BC is encoded by
non-orthogonal quantum states, $B$ cannot extract information
without disturbing the states.  This means he risks detection if
he later has to return the committed state.  At the same time, $A$ risks
detection if she sends one state and later tries to claim that she
sent the other. This suggests a strategy for
cheat sensitive BC.
The problem is to arrange for both parties to be simultaneously at
risk of cheat detection.  Standard quantum BC methods
do not work here. For example, a protocol can not be cheat
sensitive if $A$ tells $B$ at revelation what the committed
state was, since $B$ can then return a copy even if he has
disturbed the original. As we shall show, there are ways around
this difficulty.

We now describe two cheat sensitive quantum BC protocols. 
We take $\ket{0}, \ket{1}$ as orthonormal qubit
states and write $\ket{\pm} = \frac{1}{\sqrt{2}} ( \ket{0} \pm
\ket{1} )$.

{\it Protocol 1: non-relativistic CSBC}

{\it Stage 0: the prelude.}  $B$ prepares a singlet state, $|\Psi^-
\rangle_{AB}=
{\frac{1}{\sqrt{2}}}(|0\rangle_A|1\rangle_B-|1\rangle_A|0\rangle_B)$,
and sends qubit $A$ to $A$.  At certain stages of the protocol
either party may ``challenge" this singlet.  This means that the
other party must send a qubit which is supposed to be their half
of the singlet to the challenger, who can then check that the two
qubit state is indeed a singlet by measuring  the relevant
projection.  If not, there is a non-zero probability it will fail
the test.

{\it Stage 1: the commitment.}  The protocol allows a simple commitment
procedure which $A$ may use if she wishes to commit to a 
definite classical bit: to commit to $0$,
she prepares a qubit $C$ chosen randomly to be either $|0\rangle$ or $|-\rangle$,
each with probability $1/2$; to commit to $1$,
she similarly prepares either $|1\rangle$ or  $|+\rangle$.
Then she sends the $C$ qubit to $B$.

As usual in quantum BC protocols, $A$ is allowed
a more general non-classical commitment, in which she instead
prepares a state $|\psi\rangle = \sum_{r=0,1,+,-} |\alpha_r\rangle_A|r\rangle_C$,
where the unnormalised states $\ket{\alpha_r}$ are orthogonal, keeps the
$\ket{}_A$ system under her control, and sends the $\ket{C}$ qubit to $B$. 
We will show later that the probabilities for her unveiling
the classical bit are fixed by such a commitment, in the sense that they 
cannot subsequently be altered without cheating and risking detection.

{\it Stage 2: the unveiling.}  $A$ is first given the option of
challenging the singlet.  If she does and it fails the test, she
has detected cheating.  Next, whether or not she made a challenge,
she must reveal the value of the committed classical bit (but not
the qubit used to encode it).  $B$ then has the option of
challenging the singlet, if $A$ did not.    If he does and it
fails the test, he has detected cheating.

{\it Stage 3: the game.}   If either party earlier challenged the
singlet, they automatically lose the game.  If neither challenged
the singlet, they now each measure their singlet qubit in the
${|0\rangle,|1\rangle}$ basis.  $B$ sends his result to $A$.  If
hers is not opposite, she has detected cheating. If $B$ reports
the result $1$ then $A$ loses the game; if $0$ then $B$ loses.

If $A$ loses she must reveal which state was used to encode the
committed bit in the qubit $C$.  $B$ then measures $C$ to check it
is in the state $A$ claims.  If not, he has detected cheating.

If $B$ loses he must return the qubit $C$ to $A$. She makes a
measurement to check it is still in the state she originally
prepared. If not, she has detected cheating.

This completes the protocol.  Note that the cheating tests detect
only that someone -- possibly the party carrying out the test --
has cheated.  The ambiguity here is not a worry since, as usual in
mistrustful cryptography, the protocol is designed to protect
honest parties against cheats, not necessarily to protect
one cheat against another.   

A party might
choose to terminate early if they detect the other cheating. However, we
have not stipulated this, since they might choose to
continue if that seems advantageous.

{\it Proof of cheat-sensitivity:} \qquad The security proof relies on 
the following facts: 1) $B$ cannot
send anything other than a half-singlet without risking failing a
singlet challenge.  2) If A or B carry out any non-trivial
measurement on the singlet they risk failing a singlet
challenge.  If A preemptively makes her own challenge to 
avoid being challenged, she ensures she will lose the game, and this
forces her to make an honest commitment and unveiling.  
3) When A and B can no longer be challenged, they cannot
advantageously use their singlet qubit in any quantum information
processing.

As we will show, these facts imply that the singlet can be
effectively factored out from the rest of the protocol and merely
acts to provide a random ``loser'' in the game.  Neither party can
be certain they will not lose the game.  This prevents them from
cheating: $A$ may have to tell $B$ what state the qubit $C$
is in when this qubit is in $B$'s hands, and $B$ 
may have to return the qubit $C$ to $A$ in its
original state.

Clearly, at any stage, $A$ and $B$ can apply a reversible local unitary operation
to the quantum states under their control without fear of detection.
We take this as understood below,
rather than repeating the phrase ``up to a local unitary operation''
at each stage, since applying a local unitary does not {\it per se}
gain a cheat anything. 

To begin the proof, note that    
$B$ must prepare a singlet and send half of it to $A$, since
$A$ may challenge the singlet.   The following lemma shows that
once $A$ and $B$ share the singlet neither of them can carry
out nontrivial quantum operations on it.  

Lemma 1.  Suppose that A and B share a state $\ket{\Psi^-}_{AB}
\ket{\psi}_{AB}$, where $\ket{\Psi^-}_{AB} \in H^2_a \otimes H^2_b$ and
$\ket{\psi}_{AB} \in H^m_a \otimes H^n_b$, so that A's subsystem
lies in $H_A = H^2_a \otimes H^m_a$ and B's in $H_B = H^2_b
\otimes H^m_b$.  Here $H^d_a$ and $H^d_b$ denote $d$-dimensional
quantum systems initially under $A$'s and $B$'s control
respectively.  Suppose A applies a quantum measurement, defined by Kraus
operators $E_i$ corresponding to outcomes $i$, on $H_A$ and then
returns the $H^2_a$ qubit to B.  If it is the case that, for all
values of the measurement outcome $i$, B now possesses the singlet
$\ket{\Psi^-}$ in $H^2_a \otimes H^2_b$, then the Kraus operators $E_i$ must take
the form $I \otimes E'_i$, where the $E'_i$ define Kraus operators
for a quantum operation on $H^m_a$.

Proof.  $ E_i \otimes I_B \ket{\Psi^-}_{AB}
\ket{\psi}_{AB} = \ket{\Psi^-}_{AB} \ket{\psi_i}_{AB}$.  As
$E_i \otimes I_B \ket{0}_a \ket{1}_b \ket{\psi}_{AB} = \ket{0}_a
\ket{1}_b \ket{\psi_i}_{AB}$ and $E_i \otimes I_B \ket{1}_a
\ket{0}_b \ket{\psi}_{AB} = \ket{1}_a \ket{0}_b
\ket{\psi_i}_{AB}$, linearity implies that $E_i$ acts as the
identity on $H^2_a$. QED.

This implies that, unless $A$ challenges the singlet herself
before revealing her classical bit, any strategy by which she
generates the classical bit value sent to $B$ cannot involve
nontrivial operations on her singlet qubit.
Similarly, any strategy by which $B$ extracts classical information
before $A$'s unveiling cannot involve non-trivial operations on
his singlet qubit. 

Now consider an honest $A$ and dishonest $B$.  For $B$ to cheat
successfully, he must extract some information about the committed
bit before the unveiling.  We have established that he cannot
perform any operations on his singlet qubit before the classical bit is
unveiled.  Thus he must restrict his attention to qubit $C$ up to
this point.  $B$ would like to know, before the unveiling begins,
whether the committed state is $|0\rangle$ or $|-\rangle$
corresponding to a 0 or $|1\rangle$ or $|+\rangle$ corresponding
to a 1. The most general way he can extract information is to
introduce an ancilla $|P\rangle$, apply a unitary operation $U_1$, and
then measure part of the system, creating the state
\begin{equation}\label{bobsunitary}
U_1  |P\rangle|r\rangle_C = \sum_i c_r^i
|i\rangle|\eta_r^i\rangle
\end{equation}
where $r=0,1,+,-$. $B$ measures onto the states $|i\rangle$ (which
are orthonormal). For outcome $i$ he will possess the state
$|\eta_r^i\rangle$. When $A$ declares the committed bit in step
1 of the unveiling $B$ will know he has one of two states.
However, there is no deterministic algorithm for decreasing
the overlap between two states.  

We first assume $B$ makes no 
nontrivial use of his singlet qubit after $A$'s unveiling,
$B$ must thus ensure that
\begin{equation}\label{overlap}
|\langle\eta_0^i|\eta_-^i\rangle| \leq |\langle 0|-\rangle| \qquad
{\rm and} \qquad |\langle\eta_1^i|\eta_+^i\rangle| \leq |\langle
1|+\rangle|
\end{equation}
so he can send the correct state to $A$ in stage 3 if he loses
the game.  We will now see that under these conditions $B$ can
extract no information about the commitment before unveiling.
Consider applying the controlled unitary operator $U_2 =\sum_i
|i\rangle\langle i|\otimes U^i$ to the RHS of (\ref{bobsunitary}),
where
\[ U^i |\eta_0^i\rangle = |0\rangle \, ,  \qquad 
   U^i |\eta_-^i\rangle = a^i|0\rangle-b^i|1\rangle  \, ,   \]
and without loss of generality (redefining $|\eta_-^i\rangle$ 
by a phase factor if necessary) we take $a^i$ and $b^i$ to be real and positive.  From
(\ref{overlap}) it follows that $a^i\leq{\frac{1}{\sqrt{2}}}\leq b^i$.
Let $U = U_2 U_1$.  We have
\begin{eqnarray}\label{minuseqn}
\lefteqn{U |P\rangle|0\rangle = 
\Big(\sum_i c_0^i|i\rangle\Big)|0\rangle} \\
& U |P\rangle|-\rangle = \Big(\sum_i c_-^i
a^i|i\rangle\Big)|0\rangle
                              -\Big(\sum_i c_-^i
                              b^i|i\rangle\Big)|1\rangle \, .
\nonumber
\end{eqnarray}
However, we also have
\begin{equation}\label{orthog}
 U |P\rangle|-\rangle = U 
\frac{1}{\sqrt{2}}|P\rangle\Big(|0\rangle-|1\rangle\Big) 
= \frac{1}{\sqrt{2}}  \Big(\sum_i c_0^i |i\rangle\Big)|0\rangle +
|{\perp}\rangle
\end{equation} where $|{\perp}\rangle$
is orthogonal to the first term on the RHS of (\ref{orthog}).  We
can put
$|{\perp}\rangle=\alpha|A\rangle|0\rangle+\beta|B\rangle|1\rangle$
where $|\alpha|^2+|\beta|^2={\frac{1}{2}}$. Comparing
(\ref{minuseqn}) and (\ref{orthog}) we obtain
$ |\beta|^2=\sum_i |c_-^i b^i|^2\leq {1\over2} \leq \sum_i |c_-^i
a^i|^2 $. 
Since $a^i \leq b^i$, this implies 
that $a^i=b^i={1\over\sqrt{2}}$ for all $i$ for which $c_-^i \neq
0$. Hence, comparing (\ref{minuseqn}) and (\ref{orthog}) (and
since we now have $\alpha=0$) we  obtain $c_-^i=c_0^i$ for all
$i$. Also,
\begin{eqnarray}
\lefteqn{ U |P\rangle|1\rangle =U |P\rangle(|0\rangle-\sqrt{2}|-\rangle) }  \\
 & = \sum_i c_0^i |i\rangle(|0\rangle-\sqrt{2}|-\rangle)
=\sum_i c_0^i |i\rangle |1\rangle \nonumber
\end{eqnarray}
Hence, $|c_1^i|=|c_0^i|$ for all $i$, and similarly
$|c_+^i|=|c_-^i|$ for all $i$.  Thus the probability of a given
outcome $i$ cannot depend on the bit committed, and so no
information about that bit can be extracted without $B$ risking
detection.

Now we need to exclude the possibility of $B$ cheating 
without risk of detection by extracting classical information
from $A$'s commitment qubit and then carrying out some
nontrivial operation on that qubit and his singlet qubit between the unveiling 
and the game.  The most general state $A$ and $B$ may share just after
unveiling is 
$ \ket{\Psi^-}_{AB}( \ket{\alpha_1 }_A \ket{\psi_1}_B 
+ \ket{\alpha_2 }_A \ket{\psi_2}_B ) $, 
where the first state is the as yet undisturbed singlet and the second
is a state entangling two of 
$A$'s orthogonal control states $\ket{\alpha_i}$  
with two normalised states $\ket{\psi_i}$ 
resulting from $B$'s actions on the two 
states (say $\ket{0}$, $\ket{-}$)
corresponding to $A$'s unveiling c-bit.  
If $B$ has extracted useful information, we have 
$
| \braket{\psi_1}{\psi_2} | > | \braket{0}{-} | \, . 
$
$B$ must now generate a bit to tell $A$ whether or not he will challenge,
and without loss of generality we can assume he does so by applying a 
local unitary operation to create a qubit $\ket{}_c$ 
which is sent to $A$, where
$\ket{0}_c $ and $\ket{1}_c$ declare respectively
no challenge and a challenge.  Considering the constraints on $B$ implied
by the protocol, given that he wishes to avoid any risk of cheating detectably, 
we see the unitary operation must
implement a map
of the form: 
\begin{eqnarray}
\ket{ \Psi^- }_{AB} ( \ket{\alpha_1 }_A \ket{\psi_1}_B 
+ \ket{\alpha_2 }_A \ket{\psi_2}_B ) \ket{*_0}_c
\rightarrow \\ 
\ket{*_1}_{AB} ( \ket{\alpha_1 }_A \ket{0}_B + \ket{\alpha_2}_A \ket{-}_B ) \ket{1}_c 
\nonumber \\ + \ket{0}_A \ket{1}_B \ket{ *_2 }_{AB} \ket{0}_c  \nonumber \\
+ \ket{1}_A \ket{0}_B ( \ket{\alpha_1}_A \ket{0}_B + \ket{\alpha_2}_A \ket{-}_B ) 
\ket{0}_c \, , \nonumber 
\end{eqnarray}
where $\ket{*_i}$ are unspecified states.  
Considering the components
of $\ket{1}_A \ket{\alpha_i }_A $ in this equation, we see that in particular
the unitary operation must deterministically 
map states with overlap $ | \braket{\psi_1}{\psi_2} |$
to states with overlap no greater than $ | \braket{0}{-} |$.  That is,
it must deterministically decrease the overlap, which is impossible.  

Now consider an honest $B$ and a dishonest $A$.  The most general
thing $A$ can do initially is entangle the qubit $C$ with some
ancilla $A$ which she keeps.  Then the entangled state she
prepares can be written as $
|\psi\rangle = \sum_{r=0,1,+,-} |\alpha_r\rangle_A|r\rangle_C $. 
The states $|\alpha_r\rangle_A$ need not be normalized, nor
do we impose at this stage that they are orthogonal.  Further,
since the states $|r\rangle$ form an overcomplete basis, the
states $|\alpha_r\rangle$ need not be unique.  We will show that,
if $A$ is to be certain of not being caught cheating, then there
must be some way of writing the expansion above such that states
$|\alpha_r\rangle$ are indeed orthogonal (if they have non-zero
norm).  To see this consider what happens at the unveiling stage.
$A$ must announce the committed classical bit to $B$.  The most
general thing she can do is perform a unitary operation on $A$ of
the form
\begin{equation}
|\alpha_r\rangle_A \longrightarrow \sum_{k=0,1}
|\beta_{rk}\rangle_{A'}|k\rangle_{A''}
\end{equation}
and measure on $A''$ to extract a bit, $k=0,1$, which she sends to
$B$ as the committed classical bit.  At this stage $A$ may have
already challenged the singlet in which case she has lost the game
and must tell $B$ what the state of the qubit $C$ is.
Alternatively, she may not have challenged the singlet. Then $B$
may challenge the singlet.  If he declines he can immediately
measure his singlet qubit in the $|0\rangle$, $|1\rangle$ basis.
He may get a $1$ in which case $A$ loses the game and must tell
$B$ what the state of qubit $C$ is.  $A$'s singlet qubit will,
in this case, be collapsed into a definite pure state and, hence,
is of no use to $A$ in any cheat strategy. In the case where
$A$ gets $k=0$ she must be able to collapse the qubit $C$ onto
either $|0\rangle$ or $|-\rangle$ by making a measurement on $A'$
and hence we must have
\begin{equation}
\sum_{r=0,1,+,-} |\beta_{r0}\rangle |r\rangle = |u\rangle|0\rangle
+ |u^\perp\rangle |-\rangle
\end{equation}
where $\langle u|u^\perp\rangle=0$ and, similarly, for the $k=1$
case, we must have
\begin{equation}
\sum_{r=0,1,+,-} |\beta_{r1}\rangle |r\rangle = |v\rangle|0\rangle
+ |v^\perp\rangle |-\rangle
\end{equation}
where $\langle v|v^\perp\rangle=0$. This means that we can write
$\widehat{U}_A\otimes\widehat{I}_C|\psi\rangle =
|\tilde{1}\rangle_A|0\rangle_C + |\tilde{2}\rangle_A|-\rangle_C +
|\tilde{3}\rangle_A|1\rangle_C + |\tilde{4}\rangle_A|+\rangle_C
$, where $|\tilde{1}\rangle_A=|u\rangle_{A'}|0\rangle_{A''}$, etc.,
and $\widehat{U}_A$ is the unitary operator mentioned above that
$A$ may apply at the unveiling stage.  The states
$|\tilde{i}\rangle$ are orthogonal.  The probabilities that $A$
declares a $0, 1$ are given by
\begin{equation}
p_0 = \langle\tilde{1}|\tilde{1}\rangle +
\langle\tilde{2}|\tilde{2}\rangle \, , \, \, \, \, \, \, p_1 = 1 -
p_0 \, ,
\end{equation}
As the states $|\tilde{i}\rangle$ are orthogonal,
we have
\begin{equation}\label{probs}
p_0  = \Tr ( \rho_C \sigma_C ) -
\frac{1}{2} \, , \, \, \, \, \, \, p_1 = 1 - p_0 \, ,
\end{equation}
where $\rho_C = \Tr_A ( \ket{\psi} \bra{\psi} )$ and $\sigma_C =
\ket{0}_C \bra{0}_C + \ket{-}_C \bra{-}_C$.  Since these
probabilities only depend on the reduced density matrix of the
qubit $C$ which is in $B$'s hands, there is
nothing $A$ can do to alter them once she has sent this qubit to
$B$ at the commitment stage.  Hence we see that, if she is to be
certain of avoiding being detected cheating, she cannot alter the
probabilities of declaring a $0$ or a $1$.

{\it Protocol 2: relativistic CSBC}
 \qquad We now describe a simple relativistic
cheat sensitive BC protocol.  $A$ and $B$ agree two
non-orthogonal commitment states, $\ket{\psi_0}$ and
$\ket{\psi_1}$, corresponding to commitments of $0$ and $1$.
$A$ sends $B$ the state corresponding to her commitment.  When
$A$ is ready to unveil, she and $B$ set up two extra separated
sites, $A_1$ and $B_1$ relatively near the main sites $A$ and $B$
which they occupy, and $A_2$ and $B_2$ further away, with the
separations such that $ d(A,B) \approx d(A_1 , B_1 ) \approx d(A_2
, B_2 ) << d(A,A_1 ) << d(A, A_2 )$.  Each party can verify these
separations by timing the receipt of messages, so no trust is
required here\cite{akct}.  $A_2$ then reveals the committed bit to
$B_2$.  Before this 
information can reach $A,B, A_1$ and $B_1$,
these four carry out a relativistic coin tossing
protocol\cite{akct}.  If they obtain a $0$, $B$ returns the
commitment state to $A$ for testing; if a $1$, $B$ keeps it and
tests it once $B_2$ has informed him of $A_2$'s revelation.

{\it Discussion} \qquad  Quantum 
information allows us to
implement unconditionally secure cheat sensitive BC by
relatively simple protocols, which will easily be
implementable when the technology for quantum state storage
is developed.  The mechanism for cheat sensitivity used in
these protocols relies on the properties of quantum
information: classical information cannot be used in
the same way, since $A$ cannot be sure if $B$ has 
extracted information from a
classical message.  It would be interesting 
to understand how to optimise the
levels of cheat sensitivity against $A$ and $B$,
and to quantify
the effect of noise (which would tend to make our protocols
imperfectly cheat sensitive). 

Finally, our results suggest the possibility of cheat sensitive
implementations of other cryptographic tasks, such as non-relativistic quantum 
multi-party computation for general functions, for which 
unconditional security is not always attainable\cite{lo}. It 
would be interesting to
understand precisely which tasks can be implemented with cheat
sensitivity.

After submitting the first
version of this paper, we learned of independent work by
Aharonov et al.\cite{atvy}, who define and implement a related
weaker cryptographic task, quantum bit escrow.

\vskip5pt \leftline{\bf Acknowledgments}

We thank Dorit Aharonov and Amnon Ta-Shma for careful 
criticisms of earlier protocols and Jeffrey Bub, Daniel Gottesman, 
Hoi-Kwong Lo, Dominic Mayers and Rob Spekkens for helpful comments.  
We acknowledge support from Royal Society University Research Fellowships
at the Universities of Oxford and Cambridge and partial support from an
HP Bursary and the EU project PROSECCO.  AK thanks the Perimeter 
Institute for hospitality.


\end{document}